# Validation of Satellite Lifetime Predictions at Leonid Space

*Rev 2.1, authored 2026-10-21 by Scott Shambaugh*
leonidspace.com/contact

## Executive Summary

Leonid Space provides real-time satellite lifetime predictions for low Earth orbit missions. This report validates our prediction pipeline through comprehensive backtesting against 1283 satellites that deorbited between 1968 and mid-2026, demonstrating operational readiness for commercial deployment.

### Methodology

We backtested against previously deorbited non-maneuvering objects (debris, rocket bodies, and unknown), using TLE data from Space-Track and space weather records spanning six solar cycles. We simulated the final year of flight, and compared predicted deorbit dates against actuals. Our three-stage validation approach progressively removed hindsight to arrive at fully predictive operational conditions.

### Key Results

**This is the first large-scale validation of lifetime predictions using forecasted space weather** – all other studies have done so on small sets of satellites, with the benefit of space weather hindsight, or for short-horizon reentry events.

| Scenario | Ballistic Coefficient | Density Correction | Space Weather (F10.7 & Ap) | 1-Year Accuracy |
|---|---|---|---|---|
| Perfect Knowledge | Known | Known | Known | 2.4 days (0.7%) |
| Historical Atmosphere | *Predicted* | Known | Known | 9.0 days (2.5%) |
| Historical Space Weather | *Predicted* | *Predicted* | Known | 15.0 days (4.1%) |
| Fully Predictive | *Predicted* | *Predicted* | *Predicted* | 35.3 days (9.7%) |

We show that after drag coefficient estimation, **solar cycle forecasting dominates all other error sources** and higher fidelity propagators or atmosphere models are of low importance. Our approach measures the uncertainty inherent in each scenario and calibrates our Monte Carlo forecasts such that all output variance is captured in the predicted distribution and **no additional uncertainty needs to be added**.

### Implications

These results establish a performance baseline for Leonid's lifetime prediction services. The demonstrated accuracy supports mission planning, regulatory compliance, and operational decision-making for LEO satellite operators facing an increasingly dynamic atmospheric environment.

## Document History

| Revision / DOI | Prepared By | Software version | Changes | Published |
|---|---|---|---|---|
| Rev. 1.0<br>arXiv:2601.02453v1 | Scott Shambaugh | 1.0.1 | - | 2026-01-05 |
| Rev. 1.1<br>arXiv:2601.02453v2 | Scott Shambaugh | 1.4.0 | • Faster custom and Orekit propagator metrics<br>• NASA DAS accuracy comparison added | 2026-04-10 |
| Rev. 2.0<br>arXiv:2601.02453v3 | Scott Shambaugh | 2.4.0 | • Higher fidelity propagator<br>• Switch from CvM to RMSCE for calibration scores<br>• New ballistic coefficient process noise model<br>• Switch to in-house modified-MSAFE flux forecasts, and in-house Ap forecasts<br>• Density correction added<br>• Split historical into “historical atmosphere” and “historical space weather”<br>• Drop DRAMA and DAS comparisons for brevity<br>• Add 6 month and 2-year propagation comparison | 2026-09-08 |
| Rev. 2.1<br>arXiv:2601.02453v4<br>(this document) | Scott Shambaugh | 2.7.0 | • Switched from MSAFE-based to internal space weather forecasts<br>• Increased cohort from 878 to 1283<br>• RMSCE minimum selection now considers noise band | 2026-10-21 |

## Motivation

Space weather has upended the crowded environment of low Earth orbit (LEO). Densities in LEO this solar cycle have been 2-3x higher than the pre-cycle scientific consensus prediction, invalidating the lifetime assumptions in commercial business cases and national security missions. Many satellite operators have reported dramatic impacts to their constellations as a direct result of drag, including SpaceX [1], Capella Space [2], Planet Labs [3], and NASA [4]. Leonid Space recently performed an economic impact assessment of useful satellite life lost relative to the consensus prediction [5]. We assessed a conservative lower bound of $0.88 billion in impact relative to the two-sigma high forecast, and $2.77 billion relative to the nominal. This only looked at direct costs for non-propulsive payloads, and including propulsive payloads and the downstream impacts will increase this number.

Leonid Space's mission is to address this critical data gap in LEO operations. We provide real-time satellite monitoring and lifetime estimation services for satellite operators, regulators, and space domain awareness partners. This enables satellite operators to accurately predict revenue, plan production schedules, and ensure mission continuity. A live demo of our prediction pipeline is available on our website, tracking the imminent reentry of NASA's Swift Observatory [6].

The real value of a prediction is in how well it matches reality. We prove our pipeline through a comprehensive backtesting campaign of 1283 deorbited satellites, by showing that it accurately recreates their last year of flight. We start with the benefit of hindsight and use the actual TLE (Two-Line Element) ephemerides and space weather data from that year to show accuracy in a "perfect knowledge" hindcast scenario. Then through three more steps we gradually replace historical data with contemporary predictions to arrive at a "fully predictive" retrospective forecast scenario [7]. These probabilistic forecasts are scored, and we end with a measure of our prediction accuracy.

## Processing Pipeline

This paper explains the validation approach for the satellite lifetime prediction software that forms the core of Leonid's business. It is made from several components that work together in the combined pipeline shown in Figure 1: a custom high-speed orbit propagator, ballistic coefficient estimation from on-orbit TLE data, atmospheric density correction from data assimilation, custom space weather forecasts, and probabilistic Monte Carlo analysis capabilities. Our philosophy for ensuring robust predictive power is to isolate each major source of uncertainty in the underlying physics, characterize their dynamics, build forecast models for each component, and then tune a global uncertainty hyperparameter for the combined pipeline.

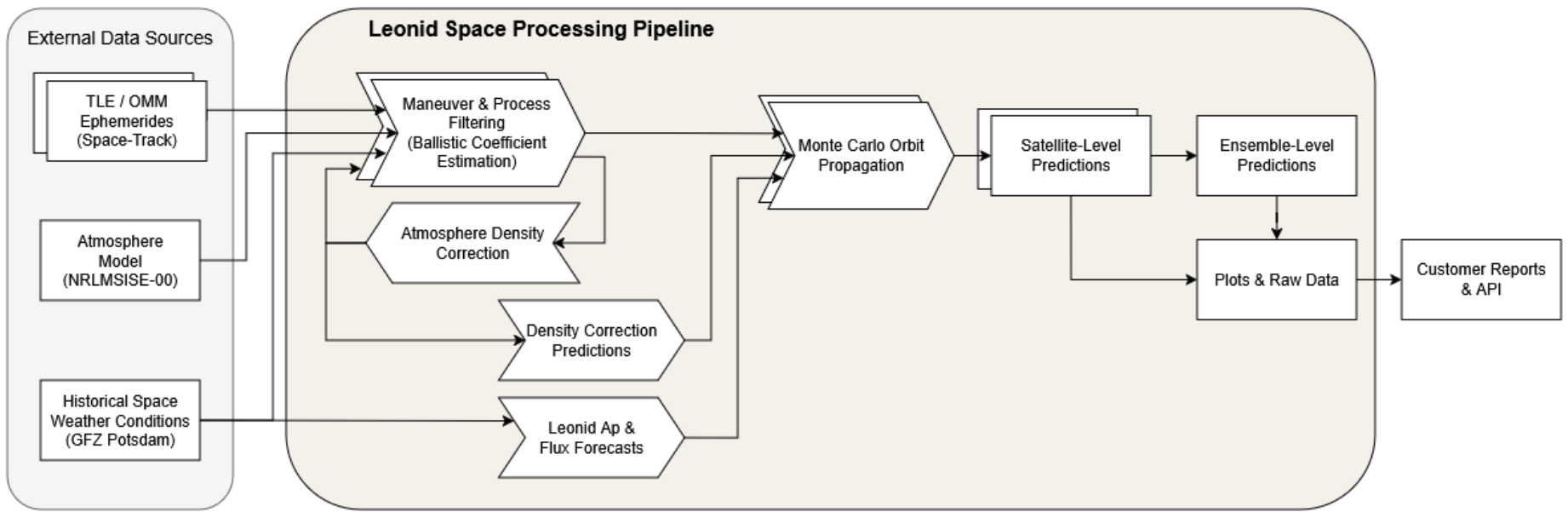

Figure 1: Diagram of Leonid's satellite lifetime prediction pipeline.

# Orbit Propagation

## Propagation Approach

Orbit propagation is performed with custom in-house software that uses a semianalytic approach based on [8], [9], and [10], which propagates mean elements for an eccentric orbit under the effects of atmospheric drag and the $J_2$ term of an oblate planet. We extended these references in two ways. First, we model a co-rotating atmosphere rather than a static one. Second, we calculate densities along each orbit using NRLMSISE-00 with specified space weather conditions. This approach strikes a balance between speed and orbit propagation accuracy. Initial conditions can be supplied manually or extracted from TLEs, and propulsive maneuvers or attitude control schemes can be modeled if known.

We additionally have extended this to a full-force model which adds in solar radiation pressure (SRP) with cylindrical eclipse shadowing, a 30x30 EGM2008 gravity field, lunisolar 3$^{rd}$ body gravity, and lunisolar tides. This greatly improves propagation accuracy over the hours-to-days periods needed for e.g. conjunction analysis. However the benefits of higher fidelity modeling are swamped by the space weather uncertainty which dominates over longer timescales, so we used the faster $J_2$ propagator described above for this analysis.

## Atmosphere Model

There are a number of thermosphere density models available to use, and good comparisons between them can be found in [11], [12], [13], and [14]. The driving use of atmosphere models in the space industry today is for conjunction screening, for which the relevant timescale is hours to days. As a result, the dimension along which most of these models compete is in capturing high-fidelity short-timescale dynamics. But they all offer similar performance over quiescent periods and longer timeframes, with the analysis in [14] showing near equivalent performance between top models over timescales as short as 48 hours. The comparisons in [15] and [16] found only a 0.1% spread between several models for reentry timing accuracy over a 28-day horizon. Even during solar storm events, density spikes occur over a period of only a few hours and so the highest-fidelity density modeling is not necessary to accurately predict lifetime when integrating over weeks to months to years. Long term bias factors between models exist, but this is automatically corrected for by using the same model for historical $C_B$ estimation and future propagation (see the discussion of debiasing via “effective density” in [14]). We use a custom in-house implementation of the NRLMSISE-00 model [17], which has been upgraded with SIMD (Single Instruction, Multiple Data) processing for faster calculations on modern CPUs.

## Propagator Validation

The Leonid $J_2$ propagator was compared against a Python-wrapped Orekit 13.1.9 implementation configured with a DSST (Draper Semianalytic Satellite Theory) propagator using $J_2$ and drag forces, a co-rotating NRLMSISE-00 atmosphere, constant space weather conditions, and a Dormand Prince 8(5,3) integrator [18] [19]. The Leonid full propagator was compared against the Orekit implementation upgraded to include 60x60 gravity, SRP, Earth albedo, relativity, lunisolar gravity, and tides. We checked a representative test case of a satellite with $C_B$=0.033 flying during constant F10.7=180 and Ap=4 space weather conditions in a 480x620 km, 51.5 deg mid-inclination orbit over a 365-day integration period. Our $J_2$ propagator took 0.082 seconds to run on a consumer laptop versus 5.55 seconds for Orekit (a ~70x speedup), and our full model took 0.107 seconds versus Orekit’s full-force 316.2 seconds (a ~3000x speedup). Taking the Orekit full force model as truth, our $J_2$ propagator final semi-major axis after one year matched to within 114 meters (0.2% of the 47.5 km drop), and our full propagator matched to within 162 meters. Notably, we are twice as fast as Orekit's SGP4 propagator, with much higher fidelity. This excellent agreement shows that our orbit-averaged quadrature approach reproduces DSST drag evolution with orders-of-magnitude faster computation, making it suitable for long-duration lifetime and Monte-Carlo analyses. The Perfect Knowledge scenario then shows that this $J_2$ propagator accurately models the real dynamics of arbitrary long-duration orbits in LEO.

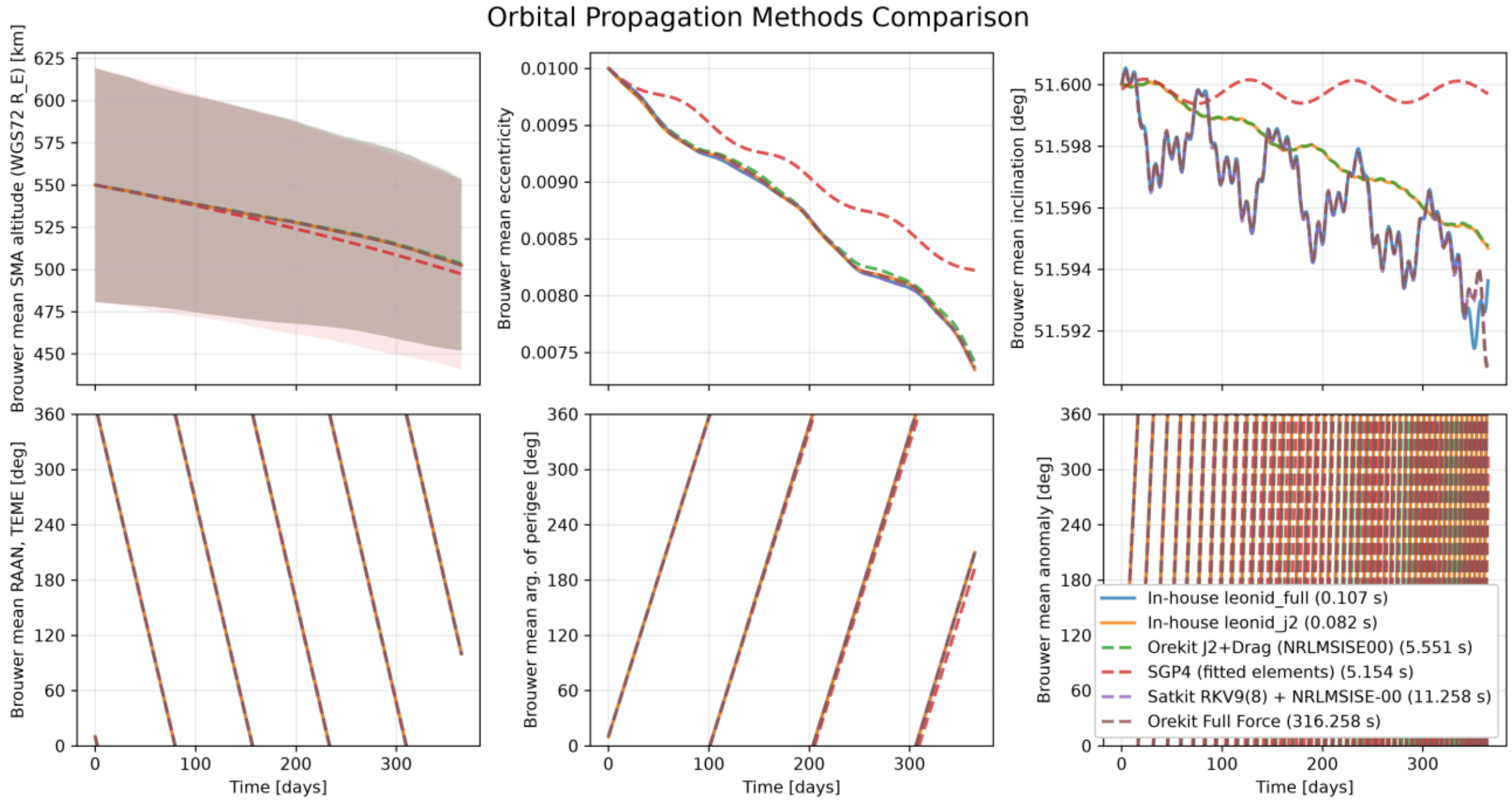

Figure 2: Leonid's custom $J_2$ and full propagators vs Orekit's DSST with $J_2$ and a co-rotating NRLMSISE-00 atmosphere, an Orekit full-force model, and Orekit's SGP4. Shaded region in SMA panel shows perigee-apogee altitude range. Angles are TLE mean element convention. Note the one-day timestep aliases mean anomaly.

## Ballistic Coefficient Estimation

The ballistic coefficient $C_B$ as a function of satellite coefficient of drag $C_D$, mass $m$, and frontal area $A$ is defined as $C_B = C_D A/m$. Note that some sources use the reciprocal of this definition.

### TLE BSTAR Parameter

The BSTAR term in TLEs notionally defines the drag component of spacecraft motion, and can be converted to $C_B$ via the equation below. A common pitfall for new astrodynamicists who are not familiar with warnings such as in [20], is that this is a fitted parameter within the SGP4 model that "soaks up" all non-modeled forces, and may bear little relation to realistic $C_B$ estimates for drag modeling. See for example section 9.4.1 in [21] for a published example of just how bad this conversion is.

$$C_B = \frac{2 * BSTAR}{R_{\text{Earth}}\rho_0} \qquad R_{Earth} = 6378.135\ km \qquad \rho_0 = 2.461 * 10^{-5} kg/m^3$$

### Instantaneous $C_B$ Extraction from TLEs

With drag as the dominant nonconservative force in LEO, we can instead extract $C_B$ from the change in mean specific orbital momentum $\bar{h}$. Between each pair of successive TLEs, the satellite state is propagated from the first to second epoch using an in-house SGP4 propagator at a small timestep (validated against *skyfield* [22]). The atmospheric density $\rho$ at each timestep is calculated using the resulting state vectors and space weather conditions, and multiplied with the squared airspeed $v_{rel}^2$ to give the dynamic pressure over the orbit. This is averaged and combined with the acceleration $f_{drag}$ extracted from the change in momentum at the given semi-major axis $a$ to estimate the ballistic coefficient. This "Energy Dissipation Rate" approach is similar to [23], [24], [25], [26], and [27].

$$\bar{h} = \overline{r \times v} \qquad f_{drag} = \frac{1}{a}\frac{d\left|\bar{h}\right|}{dt} \qquad C_B = -\frac{2f_{drag}}{\overline{\rho v_{rel}^2}}$$

### Maneuver & Process Filtering

The resulting instantaneous $C_B$ time history is a noisy signal. This is a result of both process and measurement noise sources, denoted below with (P) and (M). Thankfully for lifetime analysis, only a few of these noise sources are expected to be non-stationary processes when integrated over long timescales. Those which are generally non-stationary and may significantly bias results are marked as (P*) and (M*). For this validation campaign payloads are explicitly excluded, so the only expected non-stationary perturbation is from cross-tagging of space objects. This is rare, and we found only one instance of this in manual review of our outlier test set results (ID 3865 mis-tagged as 3863 for a 9-month period in 1971).

- (P) Changes in frontal area and coefficient of drag of the satellite as it reorients its attitude relative to the airstream.
- (P*) Changes in area or mass as the satellite deploys structures or burns propellant.
- (P*) Long-term changes in orbital operations (e.g. a switch to "high drag" mode such as has been demonstrated by SpaceX, Capella Space, and Planet).
- (P) Higher order variations in atmospheric density not captured by the NRLMSISE-00 model.
- (P) Higher order variation in atmospheric wind beyond the co-rotation assumption.
- (P) Shorter timescale and higher order variations in space weather drivers such as Ap or F10.7.
- (P*) Orbital maneuvers using on-board thrusters.
- (P) Unmodeled forces such as aerodynamic lift, Solar & Earth radiation pressure, Sun & Moon third body gravitational effects, electromagnetic drag, and higher order Earth gravitational harmonics.
- (M) TLE generation artifacts and SGP4 assumptions.
- (M) Inherent tracking uncertainty & gross measurement errors from ground-based radar systems.
- (M) Lack of cross-calibration between ground-based radar systems.
- (M*) Cross-tagging of space objects.
- (M) Calculation assumptions such as timestep quantization, integration scheme, etc.

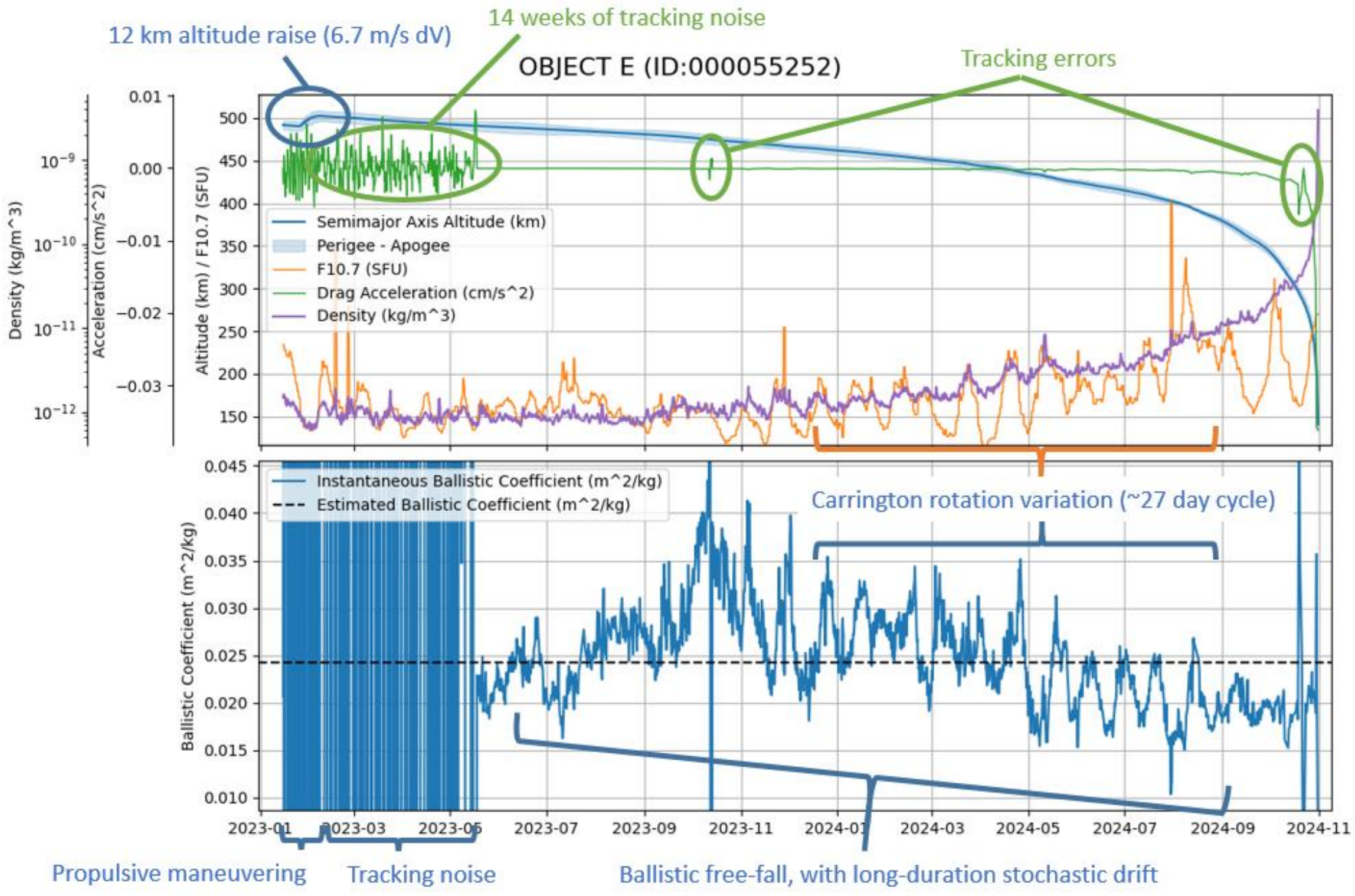

Figure 3: Processed orbital data for NORAD ID 55252 "Object E" / "Jilin-1 Mofang-02A-03". Top plot shows semimajor axis altitude (blue), F10.7 solar flux (orange), acceleration extracted from TLEs (green), and average atmospheric density seen by the spacecraft in each time interval (purple). The bottom plot shows the instantaneous $C_B$ and its estimated value for future propagation.

After maneuver detection and exclusion, determining $C_B$ from the noisy trace turns into a physically-grounded signal processing problem. Figure 3 above shows a representative trace which highlights several of the practical issues encountered: high-amplitude measurement noise which manifests both continuously over medium-duration time periods and as short duration spikes, long-duration and medium-duration (especially from the Sun's 27-day Carrington rotation period) process drift, highly variable timesteps, and negative or non-physical $C_B$ values. Leonid's approach to solving this problem is considered proprietary, and only our results will be presented here.

## Modeling Process Noise

The ballistic coefficients for the satellites are modeled as Ornstein–Uhlenbeck (OU) processes, the continuous-time equivalent of an autoregressive AR(1) model. A mean-reverting process model was chosen because a constant dispersed offset would integrate up monotonically over time, whereas the underlying physical process noise sources are bounded around a long-term mean. For example, variation in drag area is limited by the sphere of orientations that it's possible for a satellite to take.

Processes are run in log space, so that the dispersed $C_{B,t}$ at every timestep $t$ remains strictly positive and has multiplicative symmetry over its distribution. We estimate the process parameters $\mu$, $\sigma$, and $\tau$, for the OU process equation below, where $dW_t$ is incremental white noise. This process approaches a stationary distribution with log-space variance $v = \tau\sigma^2/2$. The log-space mean is set to $\mu = \ln \overline{C_B} - \tau\sigma^2/4$, so that the process stationary mean equals the nominal ballistic coefficient $\overline{C_B}$.

$$d \ln C_{B,t} = 1/\tau\left(\mu - \ln C_{B,t}\right)dt + \sigma dW_t$$

## Space Weather

### Historical Data

The NRLMSISE-00 atmosphere model uses F10.7 solar flux and the geomagnetic Ap indices as the driving space weather inputs to its density calculations. These are both obtained from the datafiles published by Geomagnetic Observatory Niemegk, GFZ Helmholtz Centre for Geosciences [28]. For data prior to 1947, we convert the SILSO SSN to F10.7 via a relationship described in [29].

### Solar Flux Forecasting

We ran a study that scored 15 F10.7 forecasts and 12 Ap forecasts from NASA's MSAFE [30] [31] [32], SWPC's 27-day, 45-day, and Solar Cycle [33] predictions, SWPC/USAF's RSGA, SIDC's Ursigram, CLS, Athens, KSO, BGS, SEPC's 3-day and 27-day forecasts, AFRL/NSO's SIFT, BoM's 3-day and Cycle predictions, NASA's Schatten, and an internal Leonid forecast. See [29] for full methodology and results.

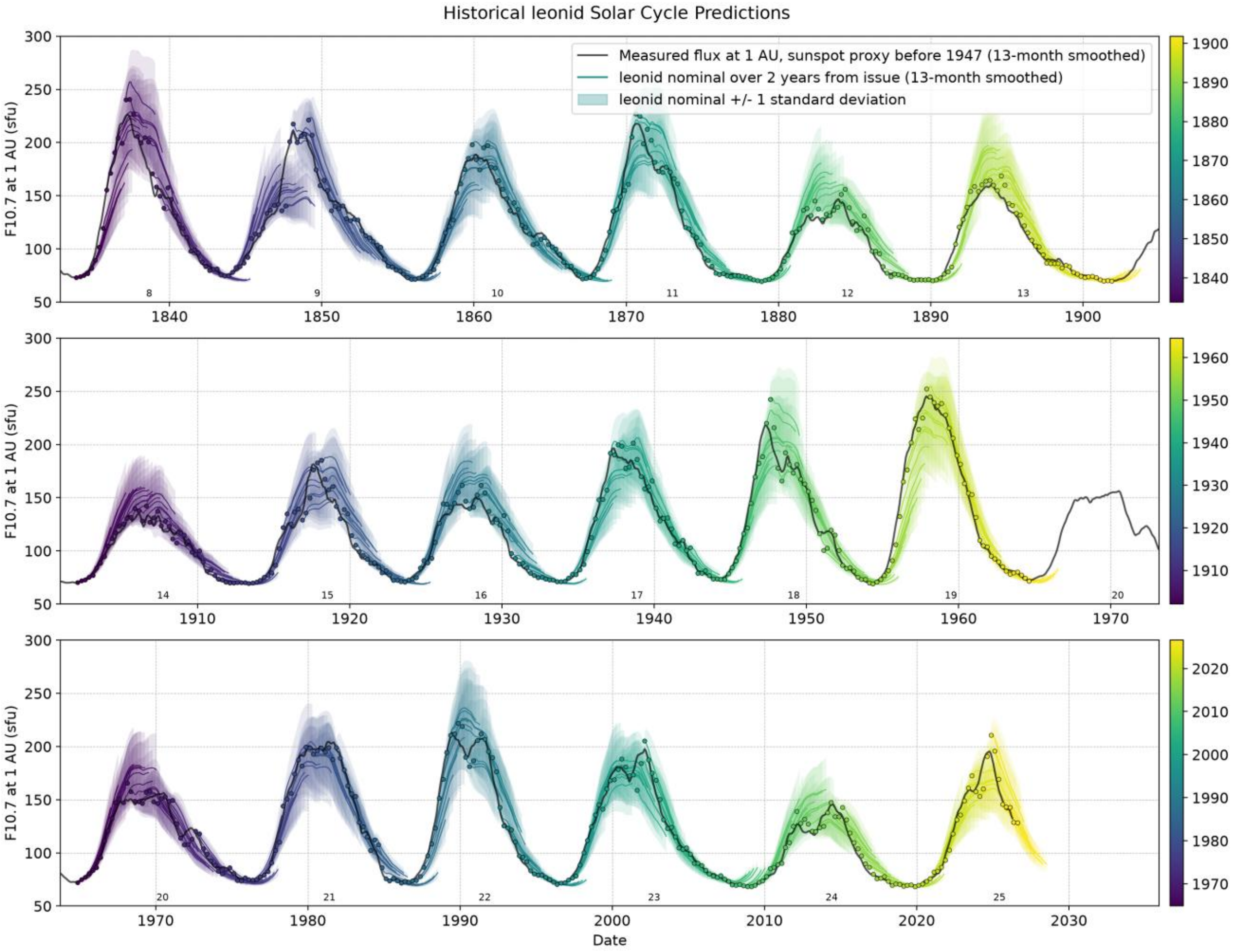

Figure 4: Leonid's internal solar flux forecast model, showing 193 years of its forecast history. These are nominally-timed forecasts with their 1-sigma uncertainty bounds, looking forward 2 years from their prediction date. Data has been smoothed to 13-month averages to better show the cycle behavior.

In v2.4.0 and prior of our software (rev 2.0 of this report), we based our solar flux forecasts on the NASA MSAFE forecast. However, several issues were apparent. Our error decomposition attributed 90% of prediction variance to this forecast, the fully predictive results had a +59 day bias due to known issues with the model [34], and it was based on 13-month smoothing of flux which caused it to lose resolution at lower timescales. This prompted the development of a unified internal forecast model, which jointly forecasts ensembles of F10.7 and Ap predictions at all timescales from days to decades. Figure 4 above shows our nominal 2-year prediction history over the past 193 years of observed flux.

## Scoring Solar Flux Prediction Models

Figure 6 below compares the CRPS, MAE, mean signed error, and RMSCE of our flux forecast model against 14 other publicly available forecast archives. These and the skill scores ($SS$) show that our model outperforms all others (including the SWPC solar cycle, 27-day [35], 45-day [36], and RSGA forecasts) at all time horizons up to 5 years, and is at parity with MSAFE after 5 years. This was scored with leave-one-out cross validation, such that we never score our model on data it was trained on. Note that the limited historical catalogue of different forecasts requires us to limit our comparison window to after 2000-10-16. Figure 5 shows the realized accuracy of a 1-day lookahead over time, and shows Leonid's forecast model to consistently outperform all others across all phases of the solar cycles since 2000, with an average one-day lookahead error of 2.87 SFU.

$$SS = 1 - \frac{\bar{C}_{leonid}}{\bar{C}_{ref}} \qquad \overline{C_s} = \frac{1}{B_s}\sum_{b=1}^{B_s}\left(\frac{1}{n_b}\sum_{i\in b} CRPS_{s,i}\right)$$

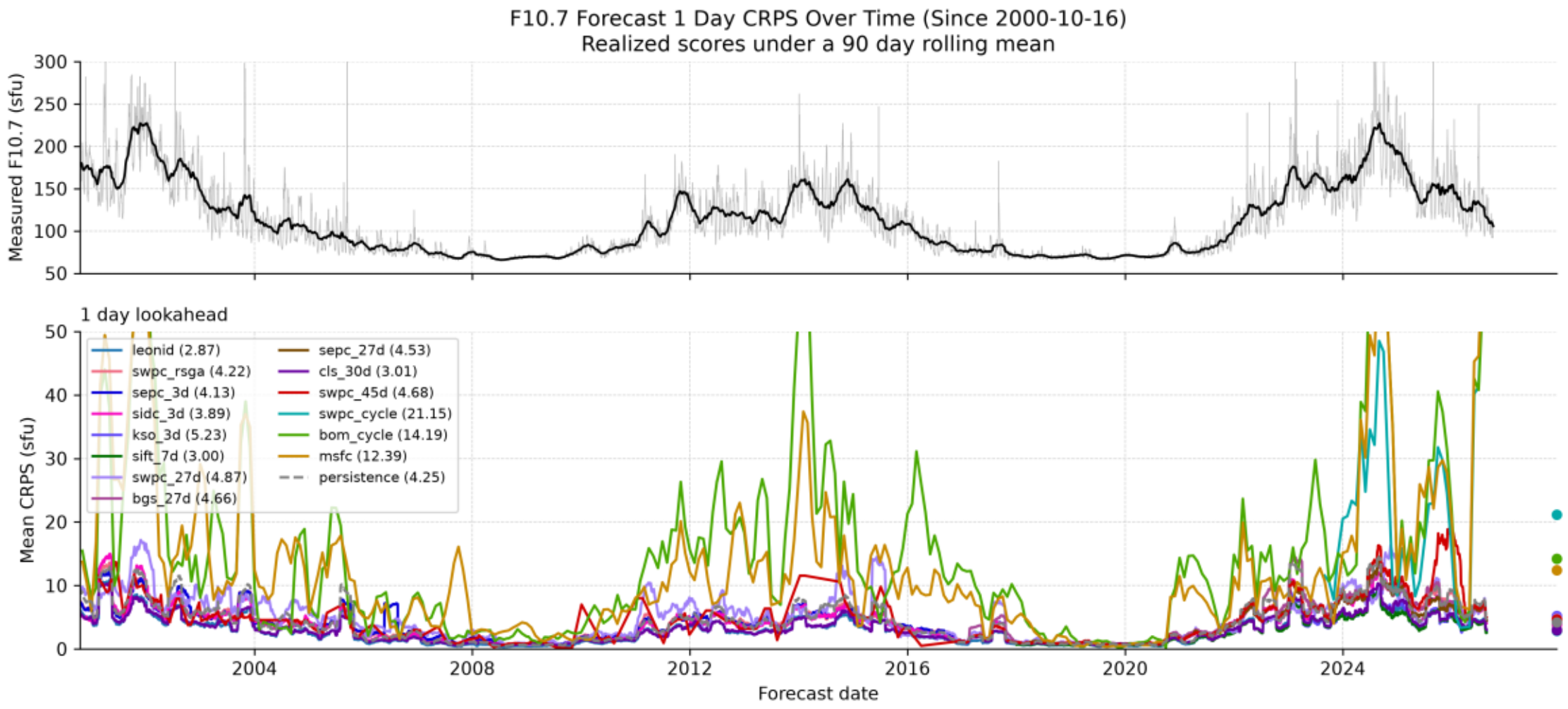

Figure 5: Comparison of the 1-day lookahead scores for different flux models over time. The CRPS (bottom) tends to scale with the intensity of the solar cycle (top). Leonid's model outperforms all other models. Dots along the right show average values – 2.87 SFU for Leonid's.

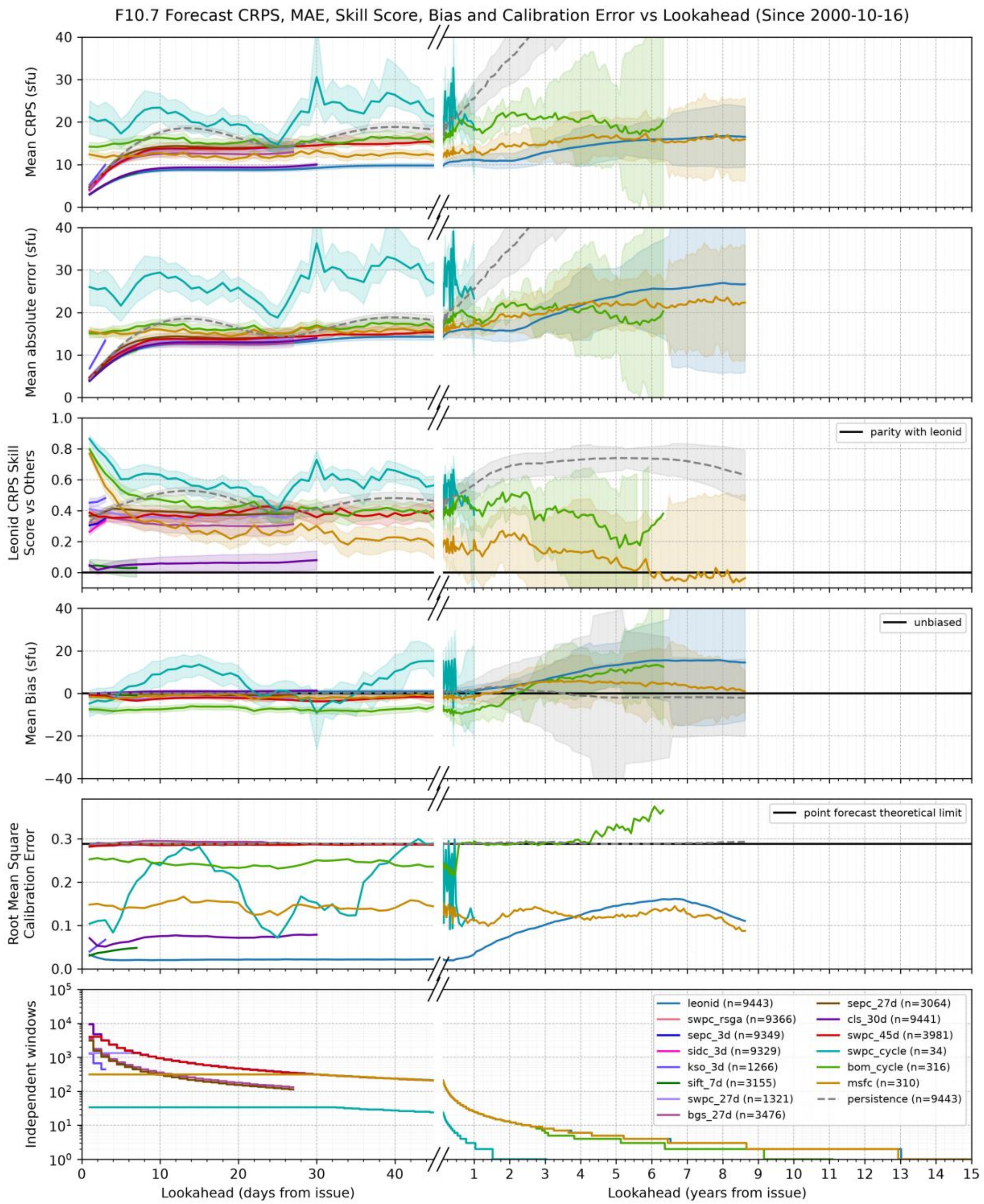

Figure 6: F10.7 forecast scores for several flux models as a function of lookahead, since 2000-10-16. Note the scale change at 45 days. From top to bottom is CRPS, MAE, Leonid's SS versus the other models, mean bias, RMSCE, and the number of independent windows used for block t-intervaling the shaded confidence intervals. We assign low confidence to data with fewer than 10 windows, and drop lookahead intervals with fewer than 3 points. Leonid's outperforms all others from 0-5 years, after which it is tied with MSAFE.

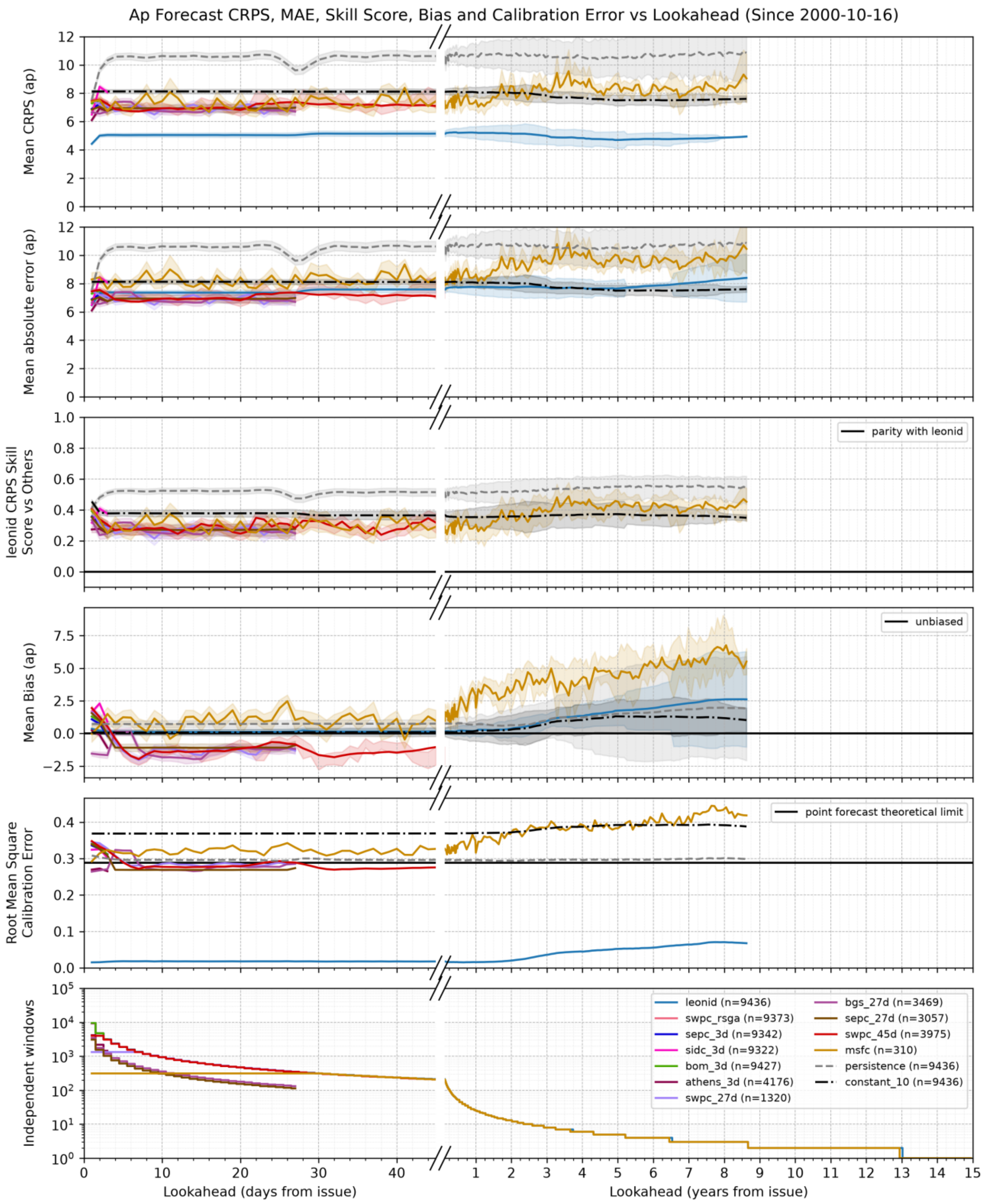

Figure 7: Comparison of geomagnetic Ap model predictive scores over time horizons up to 60 days and 15 years, showing CRPS (top) and MAE (bottom). Leonid's forecast CRPS outperforms all others over all time horizons (skill score middle). Persistence, 11-years prior, and constant $Ap = 10$ baselines are shown for comparison.

### Geomagnetic Ap Forecasting

Leonid's unified forecast model includes predictions of the geomagnetic Ap index, the details of which are proprietary but the scores of which are shown below in Figure 7 above, comparing the CRPS and MAE of our Ap forecast model against 11 other publicly available forecast archives. Our model's CRPS outperforms all others at all time horizons.

## Density Correction Through Data Assimilation

Raw ballistic coefficient estimates absorb errors in the atmospheric density model, which causes large day-to-day swings in the observed value. However these errors are highly correlated between satellites, and this correlation can be measured using a data assimilation model similar to Space Force's HASDM [37] [38] or Orion's Dragster [39], to arrive at density correction factors for a base atmosphere model. Leonid Space has built this real-time pipeline and uses it to generate a first-order global density scale factor to correct the NRLMSISE-00 model.

Our approach in version 2.0 of this report was based on a similar method as Emmert [40] [41] [42]. We gathered all debris objects flying in LEO, estimate ballistic coefficients, calculate the differences from their individual long-term averages, and correlated these per-satellite scale factors across the fleet at a 3-hour cadence. Raw ballistic coefficients were divided by the resulting global density scale factor, and those corrected ballistic coefficients are re-fed through our estimation pipeline to arrive at the values used in our propagation pipeline.

Since then, we have upgraded our density correction model to measure and account for altitude-dependent effects and 3rd order global spherical harmonics. Figure 9 shows that this process greatly stabilizes our drag estimation, reducing the average drag variance since 1970 by 70%, and over the past decade by ~90%. We use a Haar variance measure here to remove timescale-dependent windowing effects. Our implementation performs better at all epochs than the data from [42], which reduces variance over this set by 53%. The magnitude of this effect aligns with [43], which finds the first principle component of HASDM corrections to be approximately 10x the strength of any other.

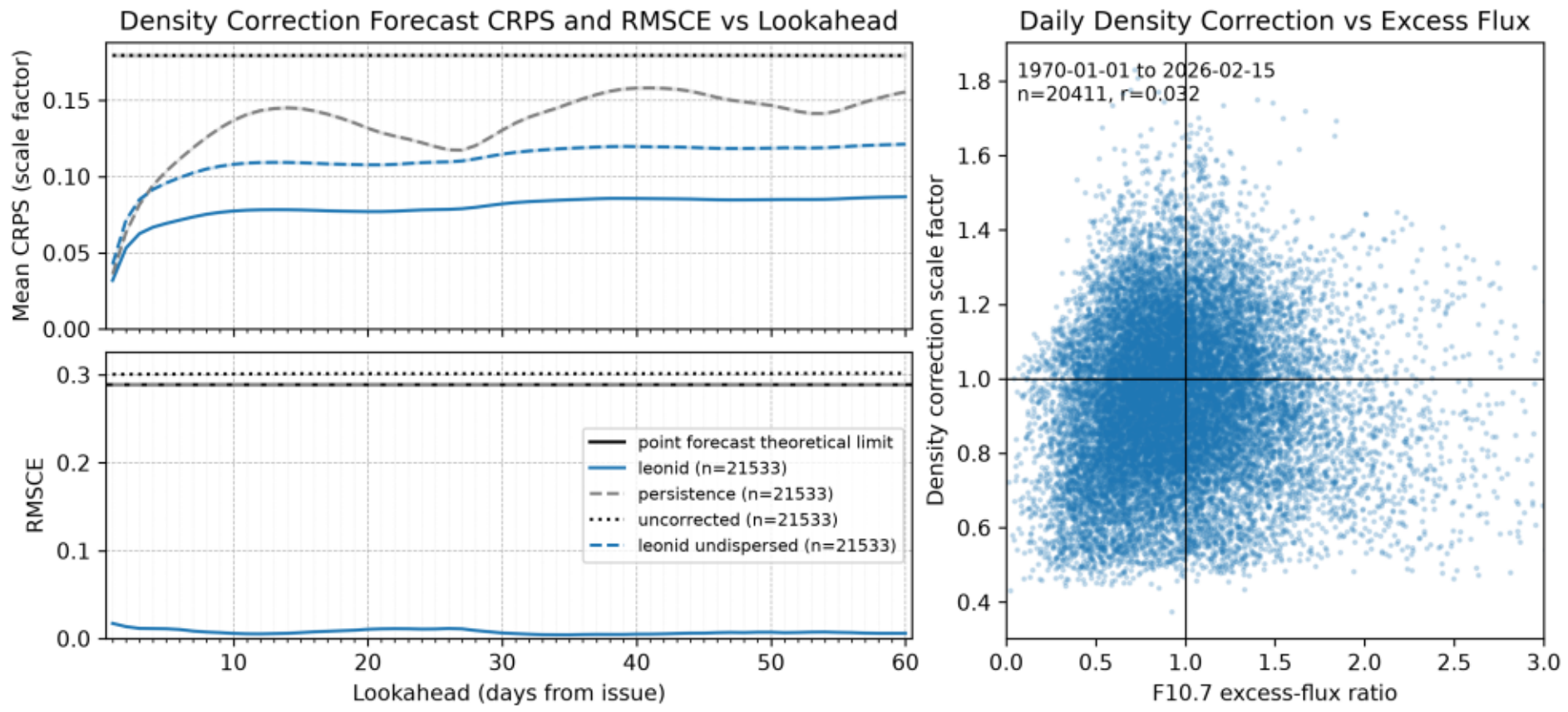

Figure 8: Scores for Leonid's density correction forecasts at different time horizons (left), showing a 6x more accurate CRPS at one day than the uncorrected density (top left), and an extremely well-calibrated <2% RMSCE over the 60-day horizon. The undispersed CRPS is equivalent to mean absolute error. Right plot shows that density correction is completely uncorrelated with solar flux residuals.

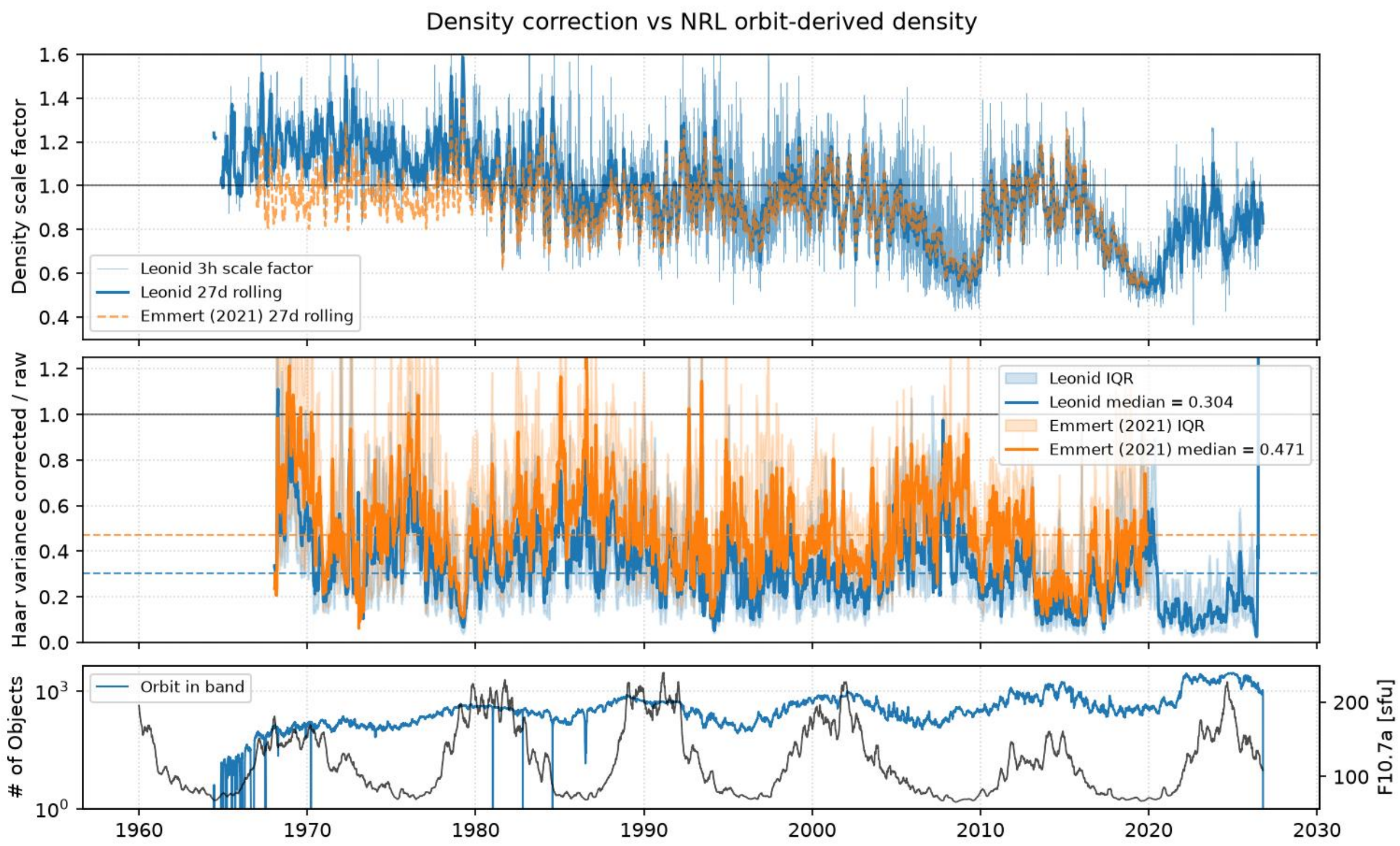

Figure 9: Leonid's global density correction scale factor versus [42] over time (top), the median 70% reduction this causes in Haar variance over this set of cohort objects during their backtest year (middle), and the number of objects in the density correction set overplotted with solar flux (bottom).

We then use these 58 years of density correction time histories to build a probabilistic predictive model of future global density correction, which is updated every 3 hours as part of our operational pipeline. The details of that model are considered proprietary and are not shared here. We back-generate and score its predicted global correction factor in Figure 8 above, comparing against an uncorrected scale factor of 1 and a persistence hold of the previous day's value. At a one-day lookahead our scale factor error of 0.032 is 6x more accurate than an uncorrected density with error 0.180. The historical intervals scored here were included when fitting our model parameters, however each historical prediction is conditioned only on observations available prior to its forecast epoch. We calculate potential bias given the long historical record, and find the maximum theoretical bias in this scoring to be negligible (<1%). We also find that the error in this model is wholly independent of the solar flux drivers – the scatter plot in Figure 8 shows the density correction factor to be uncorrelated with the 13-month flux residuals, with a linear fit of only $r = 0.032$.

## Validation Approach

### Validation Scenarios

Validation proceeds in four steps, to progressively remove the benefit of hindsight and characterize the predictive performance of each step of our processing pipeline. Table 1 summarizes these steps.

1. **Perfect Knowledge**: Simulate with actual space weather data, actual density correction values, and $C_B$ extracted from the last year of flight. This validates the end-to-end software toolchain and its ability to reproduce satellite deorbit dates versus real-world dynamics.
2. **Historical Atmosphere**: Simulate with actual space weather data, actual density scale factors, and $C_B$ estimated from the 30 days prior to simulation start. This validates extrapolating ballistic coefficients into the future.
3. **Historical Space Weather**: Simulate with actual space weather data and $C_B$ estimated from the 30 days prior to simulation start. Density scale factors are predicted rather than known. This gives the limit of how accurate propagation could be with a perfect solar flux prediction.
4. **Fully Predictive**: Simulate with forecasted space weather data, predicted density scale factors, and $C_B$ estimated from the 30 days prior to simulation start. This is analogous to a deorbit analysis performed on a spacecraft flying today, and the additional uncertainty relative to step 3 is entirely due to the unpredictability of future space weather conditions.

Table 1: Summary of the four validation scenarios, showing how we start from a perfect knowledge backtest and progressively remove hindsight to arrive at a fully predictive scenario.

| **Scenario** | **Ballistic Coefficient** | **Density Correction** | **Space Weather (F10.7 & Ap)** |
|---|---|---|---|
| Perfect Knowledge | Known | Known | Known |
| Historical Atmosphere | *Predicted* | Known | Known |
| Historical Space Weather | *Predicted* | *Predicted* | Known |
| Fully Predictive | *Predicted* | *Predicted* | *Predicted* |

### Setup

We backtested Leonid's deorbit predictions against a large dataset of past deorbit events. Using orbit data from space-track.org's TLEs [44], we predicted the date of deorbit starting from the satellite state one year prior to reentry. One year is chosen as an easily interpretable long-duration benchmark, noting that decadal scale lifetimes are driven by predictions that span multiple solar cycles [45]. For this analysis, deorbit is defined as the date when a satellite's semi-major axis falls below 300 km altitude. Below this altitude tracking data becomes sparse, so ground truth reference data is often not available. This is also an operationally-relevant threshold – at this point most objects have only a few days before they hit ~80km altitude and burn up. Predicting the precise timing or location of the reentry event is not a goal of this pipeline.

Dispersed Monte Carlo analysis is performed using the *monaco* library [46] through statistical draws from Leonid's predictive models of solar flux, Ap, and density scale factors, combined with the ballistic coefficient's randomized OU process, and run through the deterministic orbit propagator. On top of this, we add an additional global dispersion factor $g$ measured as a fraction of the mean ballistic coefficient $\mu$, such that the stationary OU process linear variance $v_{disp}$ is calculated as below. The dispersion factor $g$ represents a hyperparameter which we tune to minimize the RMSCE score in each scenario, as a characterization of the inherent uncertainty in that configuration. This setup allows us to reduce variance with a negative $g$, but that was unneeded in our results. We ran one nominal case and $n = 10$ dispersed cases for each satellite during calibration – enough to build a distribution while keeping computational runtime for many iterations over the whole ensemble reasonable. $n = 100$ was used for the final scoring.

$$v_{disp} = \max(\tau\sigma^2/2 + sign(g)\mu^2 g^2, 0)$$

## Scoring

The differences between the one-year true remaining lifetime for each satellite and the deorbit time for each case gives the prediction errors, where positive error means that the satellite deorbited earlier than predicted. The median error of all samples is calculated for each satellite as a measure of the prediction bias. The prediction accuracy is measured using an empirical continuously ranked probability score (CRPS) from the set of errors. CRPS can be thought of as a generalization of the mean absolute error (MAE) to probabilistic forecasts, and for a single non-dispersed prediction is equivalent to the absolute error.

If we have the set of predicted deorbit dates $x \in \{x_1, \dots, x_n\}$ and the true deorbit date $y$, then the CRPS value for each satellite is calculated as below. Over the whole ensemble of $m$ satellites, we look at the median of the CRPS values and median errors to score the run.

$$CRPS = \frac{1}{n}\sum_{i=1}^{n}|x_i - y| - \frac{1}{2n(n-1)}\sum_{i=1}^{n}\sum_{j=1}^{n}|x_i - x_j|$$

Additionally, we score the dispersed predictions by calculating the rank percentile of each true deorbit date within the set of predictions for each satellite. We then look at the ensemble of rank percentiles for all satellites and compare this distribution to an ideal uniform distribution, which we can plot as a histogram or a calibration curve of its empirical cumulative distribution function (ECDF). This is scored using the Root Mean Square Calibration Error (RMSCE), which measures the root mean squared difference between the ECDF and the straight-line CDF of the uniform distribution. This is the square root of the normalized Cramér–von Mises criterion Leonid used in our v1.4.0 validation report [47]. The RMSCE can be interpreted as the average percentile error of the output distribution. The equations below show the formulas, where $p_k$ is the rank percentile of the true deorbit date for the $k^{th}$ satellite and $p_{(1)} \le \cdots \le p_{(m)}$ are those values when sorted. Percentiles below 0.5 indicate that our simulations predicted a longer lifespan than reality. A perfect point forecast can only produce an even split between percentiles 0 and 1, which puts a lower bound of $RMSCE = \sqrt{1/12} \approx 0.2887$ on any forecast without associated uncertainty bounds.

$$p_k = count\{x_{k,i} \le y_k\}/n \qquad RMSCE = \sqrt{\frac{1}{12m^2} + \frac{1}{m}\sum_{k=1}^{m}\left(p_{(k)} - \frac{2k-1}{2m}\right)^2}$$

Our ultimate score for each scenario is the CRPS at the dispersion $g$ which minimizes RMSCE. This aligns with best practices for probabilistic forecasting, which says to calibrate first and then measure predictive sharpness [48] [49].

In practice, the $g$ versus RMSCE relationship is fairly flat near the minimum and the precise minimum is often a result of noise. So we run a paired bootstrap calculating RMSCE of each point against the observed minimum, and choose the value of $g$ closest to 0% for which the 95% interval of that difference includes 0. In other words, we choose the smallest dispersion with RMSCE indistinguishable from the minimum.

# Satellite Test Set

## Satellite Selection Criteria

A large test set of satellites in LEO that had deorbited due to ballistic freefall was needed. While our prediction pipeline can account for orbit control maneuvers using onboard propulsion systems, this requires information about orbit control plans and propulsive capabilities from operators that is not generally publicly available. For this reason, all "PAYLOAD" objects were excluded from this analysis. Some "UNKNOWN" objects may be active spacecraft with maneuvering capabilities, but we don't see their impact in the results.

The entire historical SATCAT catalog (~71,000 objects) was analyzed to find a set of candidate objects which could be used to validate lifetime predictions. The following filters were applied to find a set of 13,427 candidate objects:

1. OBJECT_TYPE equal to one of "DEBRIS", "ROCKET BODY", "UNKNOWN", or "OTHER". This excludes "PAYLOAD" objects.
2. Additionally not marked as a PAYLOAD in ESA's DISCOS database [50] [51].
3. RCS_SIZE not equal to "SMALL", to exclude objects with a < 0.1 $m^2$ radar cross section (RCS) in an effort to reduce measurement noise.
4. DECAY date prior to 2026-07-01T00:00:00Z.
5. Backtest start date (1 year before DECAY) on or after 1968-01-01T00:00:00Z.
6. At least 1 year and 90 days between LAUNCH and DECAY dates.

These 13,427 candidate objects were further downselected to 1,283 satellites in LEO by processing their TLEs and applying the following filters:

7. TLE data exists in space-track.org's bulk historical archive.
8. Perigee and apogee altitudes for the initial TLE between 80 and 2000 km altitude.
9. Semi-major axis altitude for the initial TLE higher than 300 km.
10. Perigee altitude for the final TLE before deorbit lower than 500 km.
11. At least 3 TLEs in the 30 days prior to backtest start.
12. At least one point in the 30 days prior to backtest start had an estimated drag force higher than estimated SRP force (drag-dominated environment).
13. NORAD ID 3863 "SL-3 DEB" appeared to be a conflation of two cross-tagged objects and was manually excluded.

## Test Set Characteristics

Figure 10 below shows the initial orbit eccentricity and inclination distribution for the test set, as well as the counts for each radar cross-section and object type categories. We see good coverage of all these parameters. With the initial orbits restricted to between 80 and 2000 km altitude, this implies a maximum eccentricity of 0.129. All 1283 satellites converged on a physically reasonable $C_B$ estimate, giving our filtering process a 100% throughput.

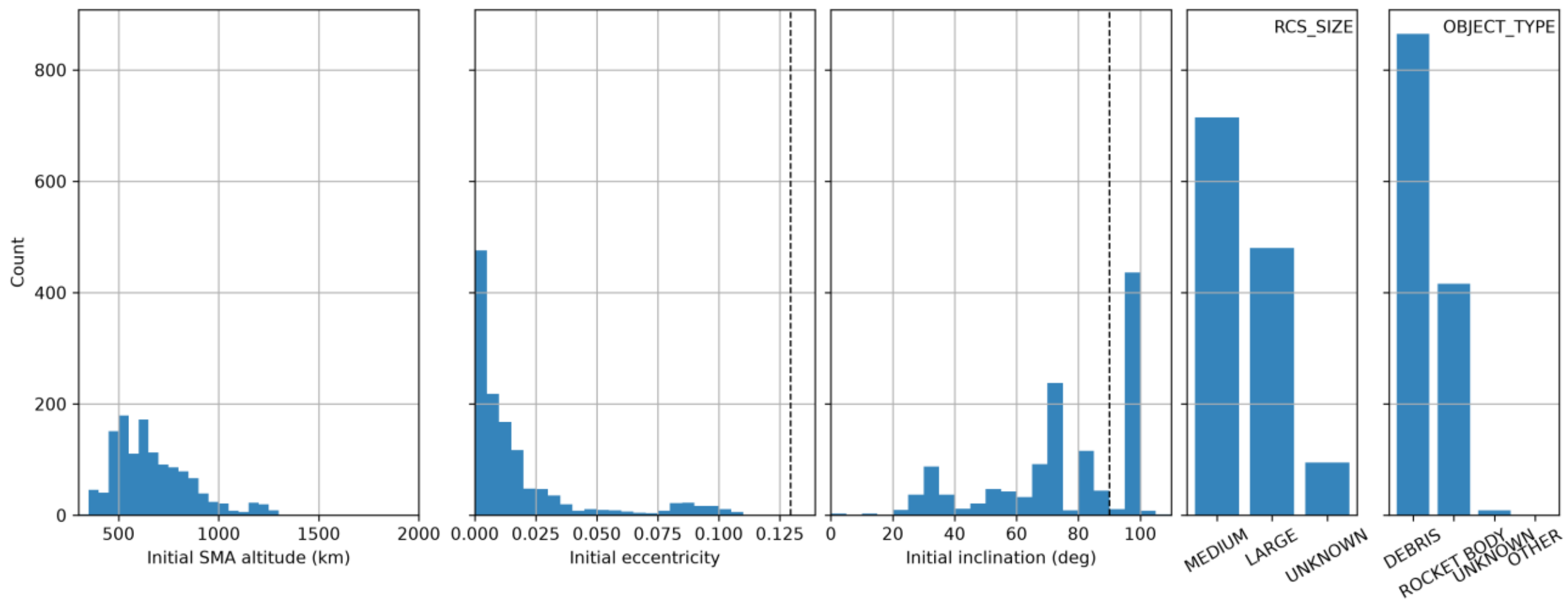

Figure 10: Characteristics of the test set of 1283 LEO spacecraft. Initial semi-major axis altitude (left), initial eccentricity with maximum possible eccentricity of 0.129 marked (middle left), initial inclination with purely polar orbit of 90 deg marked (middle), radar cross-section category (middle right), and object type category (right).

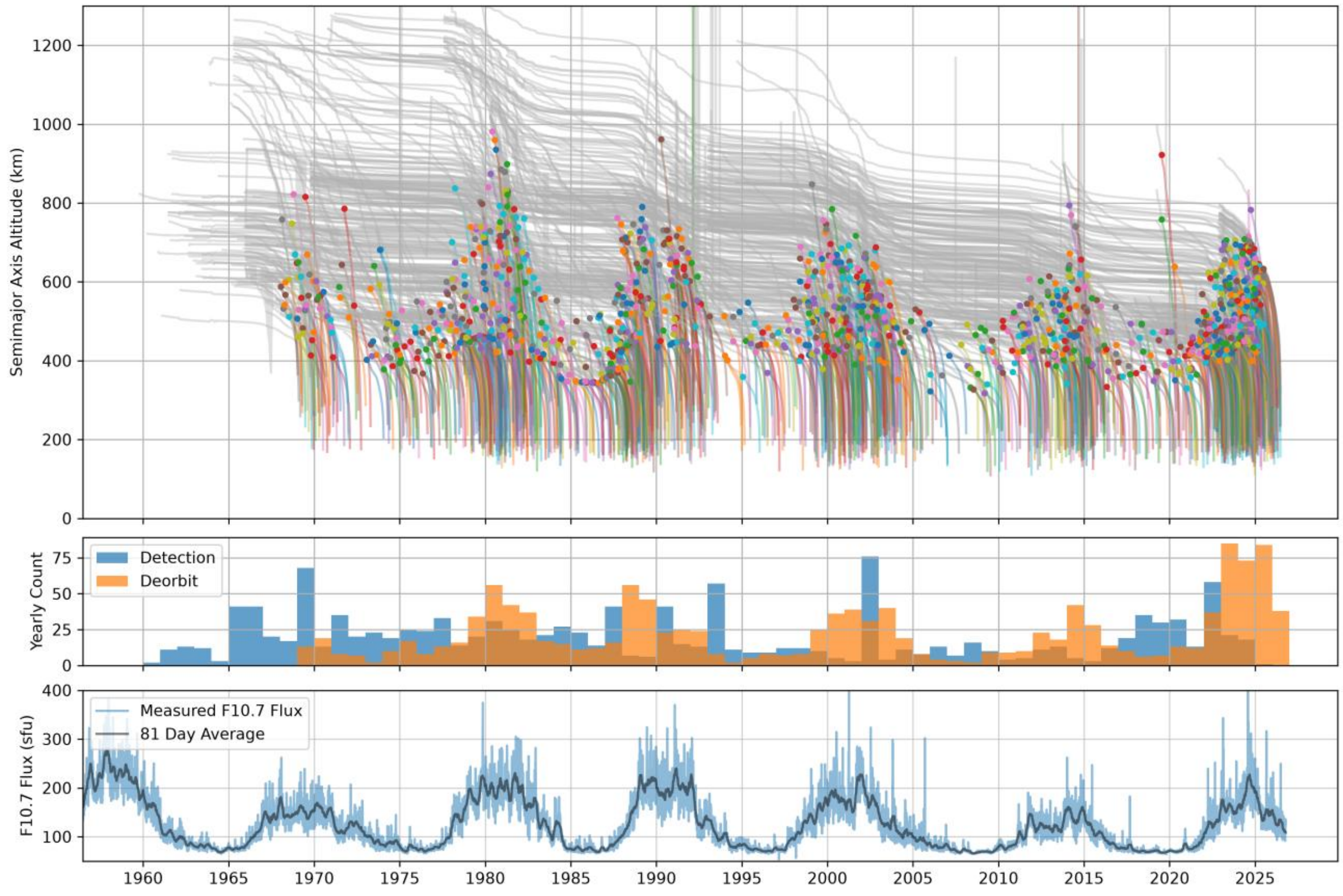

Figure 11: Semi-major axis altitude traces for the test set of 1283 LEO spacecraft (top) with the 1-year prior to deorbit colored, yearly counts of the objects being detected and deorbiting (middle), and historical solar flux (bottom). Scatter points mark our analysis start dates.

Figure 11 above shows the semi-major axis traces of all 1283 satellites in the test set. The full historical timespan of human space activity is well covered, starting soon after large-scale radar tracking began in the 1960's with the operationalization of the first Space Fence.

The effects of solar cycle are apparent with the characteristic "stairstep" pattern of steeper decay during solar maxima and flatter decay during solar minima. The resulting clustering of deorbit events during solar maxima means that the majority of the validation analysis happens during these higher-density periods of time, although the test set is large enough that there is still good coverage during the lower-density solar minima. Note also that data cuts off below 200 km – at this altitude most objects are falling fast enough to have only hours before they hit the thick portion of the atmosphere at 80 km and burn up.

# Results

## Example Propagation

Figure 12 shows a representative Perfect Knowledge example propagation for object 14824 "PALAPA B2 DEB". Starting 1 year before crossing the 300 km threshold, the estimated $C_B$ of 0.009408 m$^2$/kg results in it nominally reaching this altitude on 1990-01-12, within hours of the real object. The dispersed results have a CRPS of 2.3 days, a median error of -1.8 days, and give the true deorbit date a rank percentile of 0.55. The 95% range of dispersed cases covers a spread of ±14 days. These metrics line up well with the median ensemble results (a CRPS of 2.4 days, a median error of -0.2 days, and a 95% range of ±11 days), so this should help build intuition for what a typical deorbit prediction looks like.

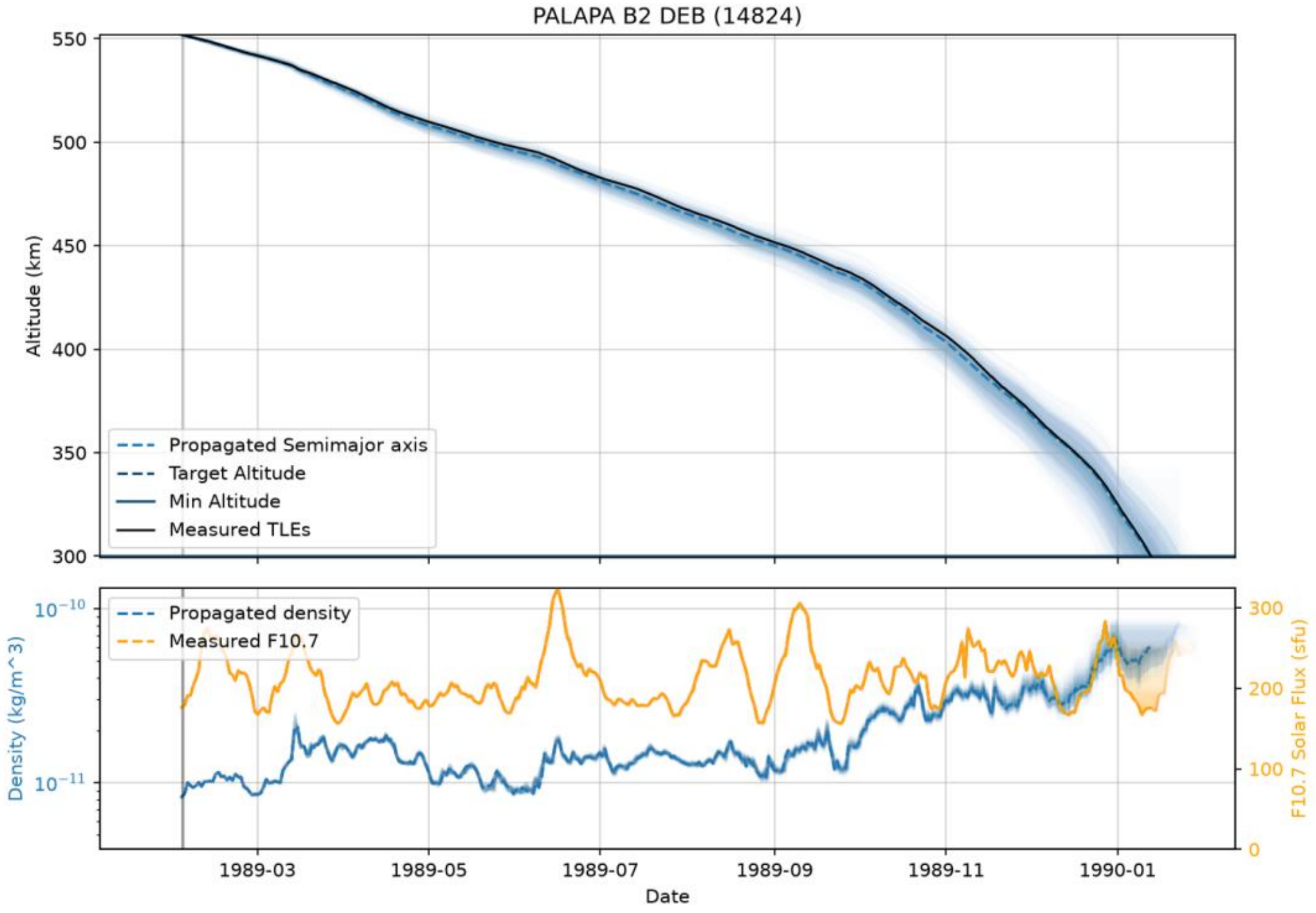

Figure 12: Propagation results for satellite ID 14824. Top plot shows the propagated semimajor axis with the dispersed cases in lighter shading, versus the true historical trajectory. Bottom plot shows the historical F10.7 flux, along with the orbit-average atmospheric density as a result of the satellite altitude and space weather conditions.

### Dispersion Calibration

Dispersing the simulation tended to reduce the RMSCE, with Figure 13 showing that as known information was removed, the optimal global drag dispersion $g$ rose from 0% for the Perfect Knowledge scenario, to 31% when $C_B$ was estimated in the Historical Atmosphere, staying at 31% when density scale factors were predicted rather than known in the Historical Space Weather. CRPS was generally flat across the low-RMSCE band, and median biases also remained low. Dispersing $g$ in the Fully Predictive scenario showed no effect on RMSCE, and $g$ = 0% was within the minimum noise band.

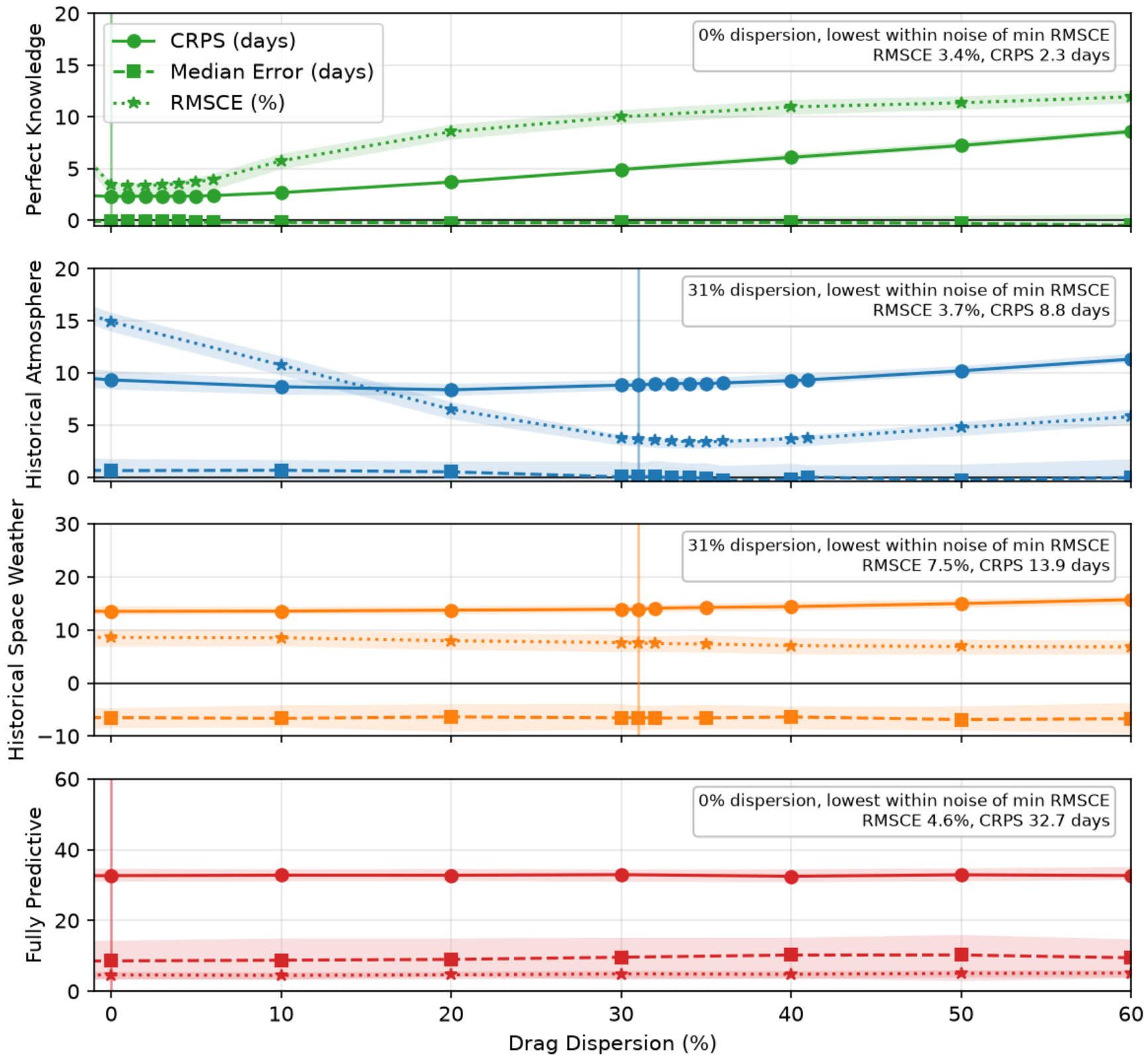

Figure 13: Median ensemble scores as a function of global drag dispersion $g$, for the Perfect Knowledge (top), Historical Atmosphere (upper middle), Historical Space Weather (lower middle), and Fully Predictive (bottom) scenarios. Shading shows the 95% confidence interval. The scores are notional given the $n = 10$ used to run the sweep, but the location of the RMSCE minimums are representative and used to determine calibrated values for $g$.

### Estimating $C_B$ from Varying Lookback Windows

For satellites currently in orbit, we only ever have historical data to work with, and must use past ephemerides to estimate $C_B$. Figure 14 below shows that while availability of TLE data worsens as the lookback window narrows, the quality of our $C_B$ estimate remains relatively stable. This is thanks to our density scale factor pipeline, and represents a significant improvement to our previous v1.4.0 pipeline, which did not use density correction and showed scores worsening monotonically from 15 to 30 days as the estimation window shortened [47].

For this analysis we use a representative 30-day lookback window – this is short enough to be actionable soon after detection of an object on-orbit, has reasonable data coverage, and allows direct comparison to the v1.4.0 results.

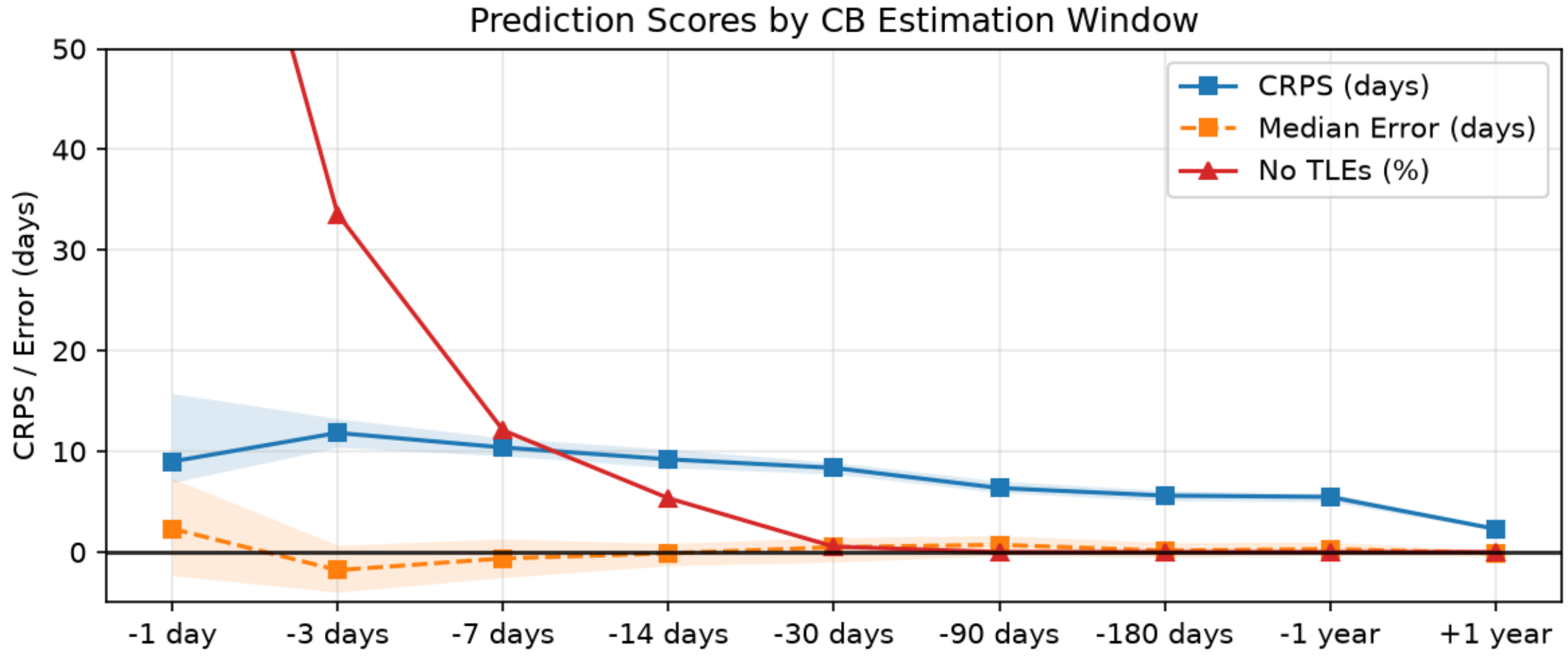

Figure 14: Ensemble CRPS and median error scores for different $C_B$ calculation lookback windows, showing that ballistic coefficient estimation is stable across lookback windows, even though TLE availability drops off. Shading shows the 95% confidence interval. The +1 year period describes $C_B$ for the Perfect Knowledge scenario. Note this cohort that was prescreened to have TLEs within 30 days.

### Error over Different Propagation Time Horizons

In the previous v1.4.0 version of our processing pipeline [47], $C_B$ was modeled as constant and the global dispersion factor $f$ was drawn from a lognormal distribution and applied to $C_B$ for each Monte Carlo case as a constant scaling factor. This constant bias would integrate up over time, meaning that $f$ was effectively tuned to the 1-year time horizon, underdispersing at shorter time horizons and overdispersing at longer ones. This was a major motivation for our move in v2.4.0 to modeling the ballistic coefficient as an OU process with stationary long-term variance, and $g$ applied as a scaling factor to that variance rather than a constant offset.

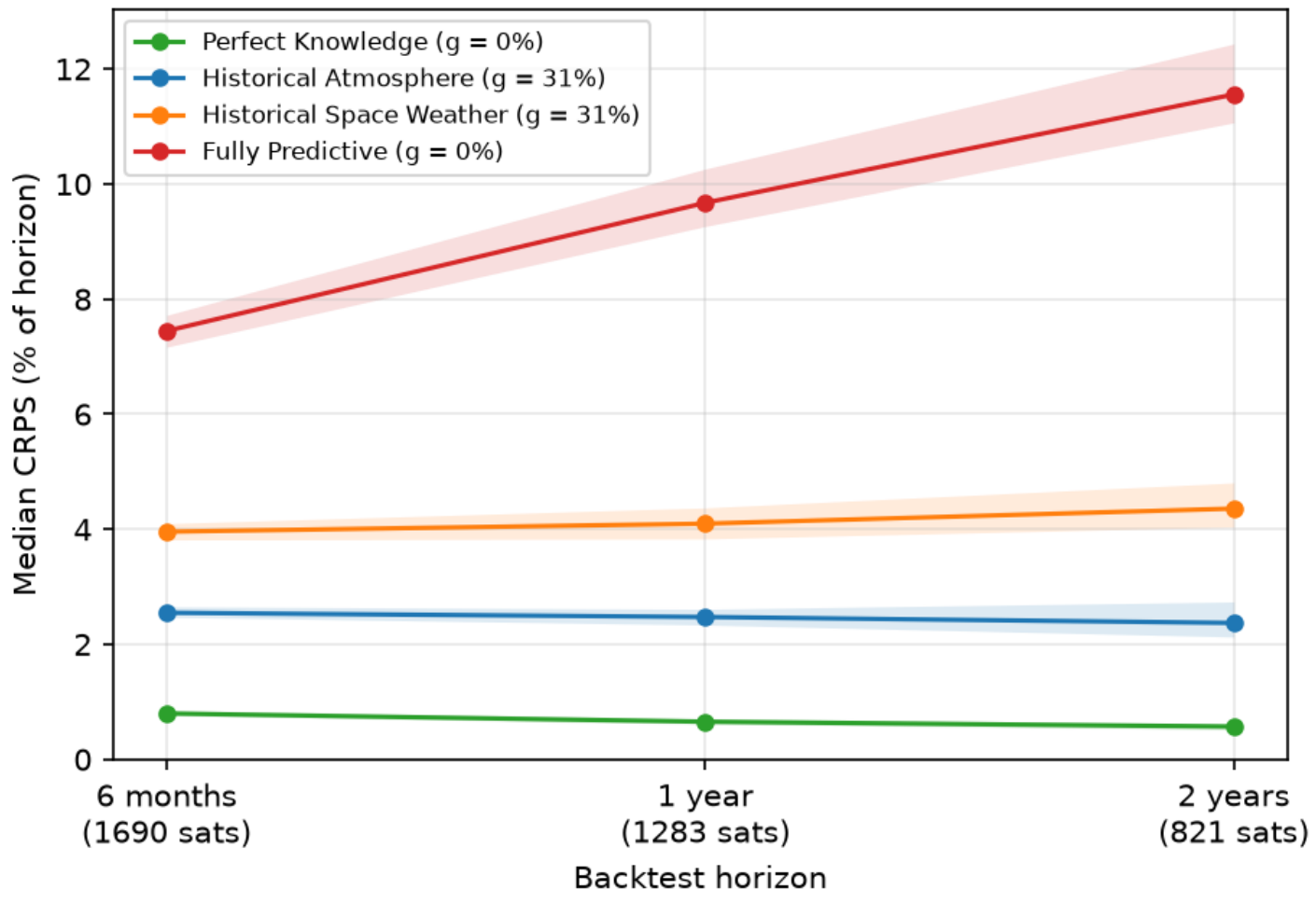

Figure 15: Median CRPS as a fraction of the time horizon for the four scenarios with global drag dispersion $g$ tuned to the 1-year horizon. Shaded regions show the bootstrapped 95% confidence interval for median CRPS. When space weather is known, relative propagation error stays flat or improves. Using predicted flux causes the relative error to grow as the backtest horizon increases from 6 months to 2 years.

To ensure that this fixed our previous overfitting to the one-year timescale, we also ran $n = 100$ propagations for 6-month and 2-year deorbit cohorts using the values of $g$ that were optimized for one year. Figure 15 shows that propagation error relative to the time horizon remains flat or improves for all scenarios with known space weather, proving that our dispersion calibration approach is timescale agnostic. The fully predictive scenario relative error grows as the time horizon increases, which again shows solar cycle forecasting to be the limiting factor for accurate deorbit predictions.

## Score Details

Figure 16 shows the distributions of CRPS and median error for the ensemble of satellite deorbit forecasts at $n = 100$, as well as the percentile histogram and calibration curves for each scenario. We can see on the top left plot that CRPS is well-clustered near zero error, with few outliers past 100 days. The top right plot shows that the median error is symmetrically distributed, with the fully predictive scenario picking up a slight +11.5 day bias.

Table 2: Summary of the one-year deorbit prediction accuracy for each scenario (left), showing the 50$^{th}$ and 90$^{th}$ percentile CRPS errors after drag calibration. These are an across-the-board improvement over Leonid Space's previous pipelines (right), although solar flux continues to dominate the fully predictive scenario. The old pipeline did not account for density corrections, so its "Historical Conditions" scenario sits near our middle two scenarios.

| | **Leonid Pipeline v2.7.0 (This work)** | | **Leonid Pipeline v2.4.0 Report Rev. 2.0** | | **Leonid Pipeline v1.4.0 Report Rev. 1.1 [47]** | |
|---|---|---|---|---|---|---|
| **Scenario** | 50% CRPS | 90% CRPS | 50% CRPS | 90% CRPS | 50% CRPS | 90% CRPS |
| Perfect Knowledge | 2.4 days | 10.3 days | 3.9 days | 13.6 days | 6.0 days | 24.0 days |
| Historical Atmosphere | 9.0 days | 37.1 days | 10.6 days | 41.8 days | 18.6 days | 80.8 days |
| Historical Space Weather | 15.0 days | 48.5 days | 15.1 days | 52.5 days | | |
| Fully Predictive | 35.3 days | 89.3 days | 44.9 days | 117.0 days | 45.5 days | 143.4 days |

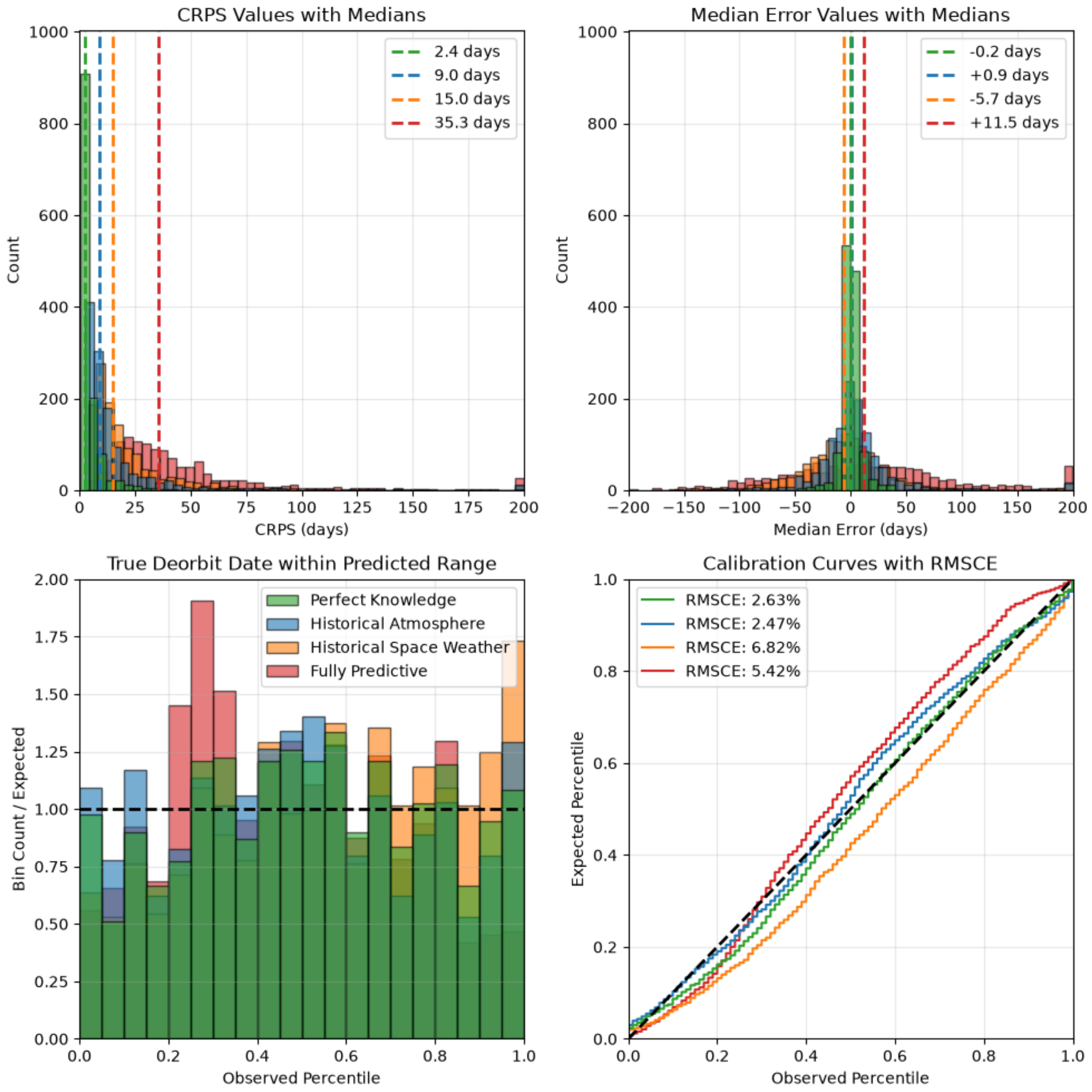

Figure 16: Ensemble prediction results for all 1283 satellites for the Perfect Knowledge (green), Historical Atmosphere (blue), and Historical Space Weather (orange) scenarios, and the Fully Predictive (red) scenario. The top row shows CRPS with the median values marked (top left) and median error with ensemble median values marked (top right). Results greater than 200 days are binned in the rightmost bar. The bottom row shows the percentile ranks of the true deorbit date within the dispersed distributions (bottom left), and the corresponding calibration curves with RMSCE scores (bottom right).

The bottom two calibration plots in Figure 16 show that the known-flux scenarios are well calibrated. These results are summarized in Table 2, and demonstrate a significant improvement of Leonid Space's previous v1.4.0 prediction pipeline. The new ballistic coefficient estimation and simulation approach, paired with the historical density correction scale factors, resulted in cutting the median error in the Perfect Knowledge scenario by a third. The tails have improved even further, with the 90$^{th}$ percentile errors being cut across the board by up to half. While the ~60 day timescale of our new predictive density correction model is too short to significantly affect one-year propagation accuracy, is has proven much more impactful over shorter timescales (not presented here). It also correctly reallocate uncertainty into the relevant physical drivers, rather than forcing the ballistic coefficient to carry the process noise from atmosphere model errors.

The move from an MSAFE-based space weather prediction model to Leonid's internal one is also demonstrated here, with a 10-day reduction in median error. While the cohort in this update has expanded due to our full timespan coverage versus MSAFE's 1999 cutoff, a matched comparison shows even better performance for Leonid's internal model. When we restrict to the 352 satellites in that fully predictive cohort we find that our 50% CRPS has improved from 44.9 to 29.9 days, and 90% CRPS from 117.0 to 66.7 days.

## Conclusion

This report validates Leonid Space's satellite lifetime prediction pipeline against 1283 deorbited satellites spanning six decades of spaceflight. Our four-stage validation methodology progressively removed hindsight to quantify accuracy under fully predictive operational conditions. The top-line results are summarized below in Table 3.

Table 3: Top-line prediction accuracy (median CRPS) results for the four validation scenarios. The percentages show the accuracy relative to the 365-day true remaining lifetime.

| Scenario | Ballistic Coefficient | Density Correction | Space Weather (F10.7 & Ap) | 1-Year Accuracy |
|---|---|---|---|---|
| Perfect Knowledge | Known | Known | Known | 2.4 days (0.7%) |
| Historical Atmosphere | *Predicted* | Known | Known | 9.0 days (2.5%) |
| Historical Space Weather | *Predicted* | *Predicted* | Known | 15.0 days (4.1%) |
| Fully Predictive | *Predicted* | *Predicted* | *Predicted* | 35.3 days (9.7%) |

If we model the step up from the Historical Space Weather 15.0 day error to the Fully Predictive 35.3 day error as a combination of normal distributions, this means that flux forecasts are contributing 75% of the variance in our deorbit predictions. This dominates all other sources of error. But it also establishes a new framework for scoring solar cycle forecasts, by grounding forecast accuracy in operationally-relevant satellite lifetime predictions. This lays the groundwork for future development of improved solar cycle forecasts.

To our knowledge, this is the first comprehensive backtesting of long-term satellite lifetime predictions that reports quantitative results for the fully predictive case. We qualify this by requiring comparison to real orbit data, setting a modest threshold of "comprehensive" as $n > 100$ satellites, and defining "long duration" prediction as >30 days. This screens out a large number of studies which parametrically evaluate models, focus on small cohorts of selected satellites, or look at short duration conjunction assessments or reentry location predictions. Notable examples in this camp include [16], [52], [53], [54], [55], [56], [57], [58], [59], [60], and [61]. The studies in [21], [45], [62], [63], [64], and [65] all report hindtesting results that fit the qualifying criteria, but only for known historical space weather conditions. [65] looks at a fully predictive scenario, but unfortunately only reports which of two space weather forecasts gives relatively better results rather than quantitative accuracy. This means that the results for Leonid's analysis pipeline are uniquely valuable in an operational context where the future is not yet known.

The results here are based on TLE data and non-payload objects that are tumbling without attitude control. This is a noisier test set than stable payload platforms. When working with payload operators, Leonid's analysis can use higher-fidelity onboard GPS telemetry and include orbit control plans, propulsion system characteristics, and attitude control schemes, which will eliminate process and measurement noise sources. For those operators we expect that our prediction accuracy will only improve on the results here.